\documentclass[letter]{aa}  
\usepackage{graphicx}
\usepackage{txfonts}
\usepackage{natbib}
\bibpunct{(}{)}{;}{a}{}{,} 
\usepackage[colorlinks=true,linkcolor=blue,citecolor=blue,urlcolor=blue]{hyperref}
\usepackage{caption}
\usepackage{booktabs}  
\usepackage{siunitx} 
\usepackage{amsmath} 
\usepackage{breqn}
\usepackage{amssymb}
\usepackage{graphicx}
\usepackage{float} 
\usepackage{wrapfig}

\defcitealias{2012diaz}{D12}
\defcitealias{2021florian}{FN21}
\defcitealias{2022florian}{FN22}
\defcitealias{2021zivick}{Z21}
\defcitealias{2021GaiaMC}{G21}
\defcitealias{2025nakano}{N25}
\defcitealias{2025dhanush}{D25}
\defcitealias{2025vijayasree}{V25}

\begin{document} 

   \title{The VMC survey} 
   \subtitle{LV. The coherent expansion of the SMC}
   
   \author{S. Vijayasree,\inst{1,2}
   F. Niederhofer,\inst{2}
   M.-R. L. Cioni,\inst{2}
   J. Th. van Loon, \inst{3}
   K. Bekki, \inst{4}
   R. de Grijs, \inst{5,6,7}
   S. Subramanian, \inst{2,8}
   N. Kacharov, \inst{2}
   A. O. Omkumar, \inst{1,2}
   L. R. Cullinane, \inst{2}
   V. D. Ivanov \inst{9}
    }
    \authorrunning{Vijayasree et al.}
    
   \institute{Institut für Physik und Astronomie, Universität Potsdam, Haus 28, Karl-Liebknecht-Str. 24/25, D-14476 Golm (Potsdam), Germany
   \and
   Leibniz-Institut für Astrophysik Potsdam, An der Sternwarte 16, D-14482 Potsdam, Germany
   \and
   Lennard-Jones Laboratories, Keele University, Keele ST5 5BG, United Kingdom
   \and
   ICRAR, M468, The University of Western Australia, 35 Stirling Hwy, Crawley, Western Australia 6009, Australia
   \and
   School of Mathematical and Physical Sciences, Macquarie University, Balaclava Road, Sydney, NSW 2109, Australia
   \and
   Astrophysics and Space Technologies Research Centre, Macquarie University, Balaclava Road, Sydney, NSW 2109, Australia
   \and
   International Space Science Institute--Beijing, 1 Nanertiao, Zhongguancun, Hai Dian District, Beijing 100190, China
   \and
   Indian Institute of Astrophysics, Koramangala II Block, Bangalore-560034, India
   \and
   European Southern Observatory, Karl-Schwarzschild-Strasse 2, D-85748 Garching bei München, Germany
   }

  \date{Received ; accepted }

\abstract
{The Small Magellanic Cloud (SMC) exhibits significant kinematic disequilibrium due to interactions with the Large Magellanic Cloud (LMC).}
{Here, we investigate the two-dimensional stellar kinematics of the SMC to understand the dynamical effects of these interactions by exploiting the increased time baseline of 6$-$11 years from the VISTA Survey of the Magellanic Clouds (VMC) data release 7.}
{We derive proper motions with a threefold improvement in precision compared to previous studies based on VMC data. We used a geometric framework accounting for perspective effects from line-of-sight motion to model the systemic motion across the SMC and construct a residual proper motion map. We further introduce an anisotropic linear velocity gradient model to quantify the stretching of the galaxy.}
{For the first time across all stellar populations, the residual proper motion map reveals expansion along the south-east and north-west directions, consistent with LMC-induced tidal forces, detectable even in the central regions. The gradient-corrected residuals show predominantly radial motions towards the SMC centre with no evidence of rotation. Velocity maps for different stellar populations, without assuming a rotating-disk model, reveal a coherent northward motion away from the centre exclusively in older red giant branch stars, interpreted as a kinematic signature of a past (>2 Gyr ago) interaction.} 
{This study highlights the inadequacy of simple rotating-disk models in capturing the internal kinematics of the galaxy.}


\maketitle

\section{Introduction}

The Small Magellanic Cloud (SMC) and the Large Magellanic Cloud (LMC) are one of the closest pairs of satellite galaxies of the Milky Way (MW). Located at a distance of $\sim$ 62 kpc \citep{2015degrijs, 2020graczyk} and  exhibiting a large line-of-sight depth of up to $\sim$ 30 kpc depending on the stellar population considered \citep[hereafter D25]{2025dhanush}, the SMC is classified as a dwarf irregular galaxy (dIrr; \citealt{1959vandenbergh}) whose morphology and kinematics are strongly shaped by interactions with the LMC. The dynamical effects of these interactions are encoded in the motions of stars, which motivated a detailed study of the stellar kinematics using spatially resolved velocity maps, given its relative proximity. With the advent of highly precise proper motion measurements from the \textit{Hubble} Space Telescope (HST), the VISTA survey of the  Magellanic Clouds \citep[VMC;][]{2011cioni} and the \textit{Gaia} mission \citep{2016GaiaMission}, several spatially resolved stellar proper motion studies have been conducted. 

Early HST measurements of the SMC centre-of-mass (COM) proper motion revealed a large relative velocity between the Magellanic Clouds of 105$\pm$42 km s$^{-1}$ \citep{2006kallivayalilsmc}, constraining their infall into the MW halo \citep{2007besla}. Subsequent seven-year baseline measurements refined the COM motion \citep{2013kallivayalil}, while expanded HST coverage revealed a dynamically disturbed internal kinematic structure consistent with a direct LMC–SMC collision $\sim$ 150 Myr ago \citep{2018zivick}. The first homogeneous, large-scale proper motion maps of the Magellanic Clouds were produced with \textit{Gaia} DR2 \citep{2018gaia}, while \textit{Gaia} Early Data Release 3 \citep[EDR3;][hereafter G21]{2021GaiaMC} delivered substantially improved precision and cleaner SMC member selection. Analyses of \textit{Gaia} EDR3 data showed that the internal kinematics of the SMC are population dependent: older stars are largely dispersion dominated and younger populations in the inner regions exhibit rotation-like motions (\citealt[hereafter Z21]{2021zivick}; \citealt{2021piatti}). More recent \textit{Gaia} DR3 studies, using multiple stellar tracers, further revealed residual motions and kinematic substructure indicative of ongoing tidal disturbances (\citetalias{2025dhanush}; \citealt[hereafter N25]{2025nakano}). Kinematic studies of the SMC using the VMC dataset (\citealt{2016cioni,2018florian2,2021florian}, hereafter FN21) with a time baseline of $\sim$3 years derived a COM proper motion comparable in precision to space-based measurements. 

For this work we expanded the proper motion study of the SMC, using multi-epoch near-infrared observations from the VMC survey, with a total time baseline of $\sim$ 11 years \citep{2025cioni}, which shows a significant improvement over previous work. We adopted the methodology validated for VMC-based proper motions (\citetalias{2021florian}; \citealt{2022florian}; \citealt[hereafter V25]{2025vijayasree}) to produce a comprehensive residual velocity map by subtracting the systemic motion using a simple geometric model. The expanded temporal baseline allowed us to better resolve the tidal stretching of the galaxy in opposite directions with greater accuracy. For the first time, we tried to model this tidal stretching motion using a diverging linear velocity gradient. By exploiting multi-band VMC photometry, we also explored multiple population-dependent variations in these residual motions, without invoking a rotating-disk model, taking into account the complex morphology and kinematics of the SMC.

\section{Data} 

The VMC is a near-infrared survey of the Magellanic Clouds conducted in the \textit{Y}, \textit{J}, and \textit{K}$_{\mathrm{s}}$ bands. It uses the VISTA Infrared Camera \citep[VIRCAM;][]{2006dalton} mounted on the Visible and Infrared Survey Telescope for Astronomy \citep[VISTA;][]{2010emerson} and has a field of view of 1.65 deg in diameter. VIRCAM is equipped with 16 detectors arranged in a 4$\times$4 grid with gaps between the detectors. To observe a contiguous area of the sky, the telescope uses large dither patterns, having 90$\%$ and 42.5$\%$ of detector size offsets in the \textit{x} and \textit{y} directions, respectively; these individual images are called pawprints. A single VMC tile is made of six such pawprints. The survey ran for approximately nine years, from November 2009 to October 2018. To extend the temporal coverage, the programme was later expanded by almost four years; we refer to this extended phase as VMCExt.

For the SMC, the survey consists of 27 tiles covering a total area of 42 deg$^2$. The dataset used in this study was observed between October 2010 and December 2022, resulting in a time baseline of roughly 11 years; four times longer than that available for the previous study using VMC data \citepalias{2021florian}. Each tile has a total of 14 epochs of $K_{\mathrm{s}}$-band observations, comprising 13 from the original VMC survey and one additional epoch from the VMCExt. Not all tiles, however, benefited from the extended baseline. In particular, for tile SMC 5$\_$4, the VMCExt epoch was obtained through a separate programme that employed a different jitter pattern than the main survey, and for tile SMC 5$\_$2, the detectors were swapped during observation in the VMCExt epoch. Therefore, these tiles were excluded from obtaining the systemic motion of the SMC, although they were included in the velocity maps to avoid gaps (see Fig. \ref{fig:vel_map}). Additionally, in the original VMC survey, there was a gap between tiles SMC 5$\_$3 and 5$\_$4, which was later filled with observations of the region designated as the SMC gap, taken from September 2017 to December 2022. The details of the 28 SMC tiles, including the gap tile, are provided in Table \ref{tab:Tiles}, along with the time baselines for the original VMC survey and the VMCExt.

The individual pawprint images were downloaded from the VISTA Science Archive\footnote{\url{http://vsa.roe.ac.uk}} \citep[VSA;][]{2012cross}, which were already pre-processed by the Cambridge Astronomy Survey Unit (CASU) through the VISTA Data Flow System \citep[VDFS v1.5;][]{2004Irwin,2018Gonz}. For the proper motion calculations, stellar centroids were determined by performing point spread function (PSF) photometry on the detector-level images using a dedicated photometry pipeline. Comprehensive discussions of the procedure were presented in \citet{2015rubele}, \citetalias{2021florian}, and \citetalias{2025vijayasree}.

Proper motions were derived following the methodology of \citetalias{2022florian} and \citetalias{2025vijayasree}. Briefly, PSF-based catalogues for each tile containing stellar centroids were split into stellar sources and background galaxies using predefined selection criteria. 
All epochs were transformed to a common reference frame using likely SMC stars selected from \textit{Gaia} EDR3 and near-infrared 
colour-magnitude diagrams (CMDs; see Appendix~\ref{app:cmd}), with the epoch of best seeing adopted as the reference. This procedure defines a relative frame tied to the mean SMC motion, such that the reference population is stationary on average. Proper motions were obtained by linear least-squares fits to the stellar centroid positions in the \textit{x} and \textit{y} directions as a function of epoch. Foreground contaminants were removed by cross-matching with \textit{Gaia} EDR3 and applying additional CMD selections to isolate probable SMC members. The relative proper motions were then placed on an absolute (heliocentric) scale, yielding a final catalogue of 1,461,922 probable SMC sources, corresponding to 759,161 unique sources due to the pawprint overlap.

Following standard conventions, we define the proper motion components as $\mu_W = -\mu_\alpha \cos(\delta)$ for right ascension (RA, the proper motion in the western direction) and $\mu_N = \mu_\delta$ for declination (Dec, the proper motion in the northern direction). The median proper motion of our sample is $\mu_W = -0.693$ mas yr$^{-1}$ and $\mu_N = -1.230$ mas yr$^{-1}$, with standard deviations of 1.997 and 2.051 mas yr$^{-1}$, respectively, indicating an overall motion towards the south--south-east direction. These uncertainties are three times smaller than those of \citetalias{2021florian} based on VMC data, owing to the longer time baseline employed in this work. The final proper motion catalogue was also tested for confusion in both the outer (\textit{r} > $\sim$ 2 kpc) and the more crowded inner (\textit{r} $\leq$ $\sim$ 2 kpc) regions and is generally not confusion-limited for the adopted spatial binning (see Appendix~\ref{app:confusion_limit}).

\section{Stellar kinematics from proper motion maps}

\subsection{Kinematic modelling}

The improved proper motion measurements allowed us to calibrate the kinematic parameters for the SMC with high precision. The kinematic modelling was performed following the geometric framework of \citetalias{2021florian}. A detailed description of the model is given in Appendix \ref{app:model}. 

\begin{figure*}
    \centering
    \includegraphics[width=\columnwidth]{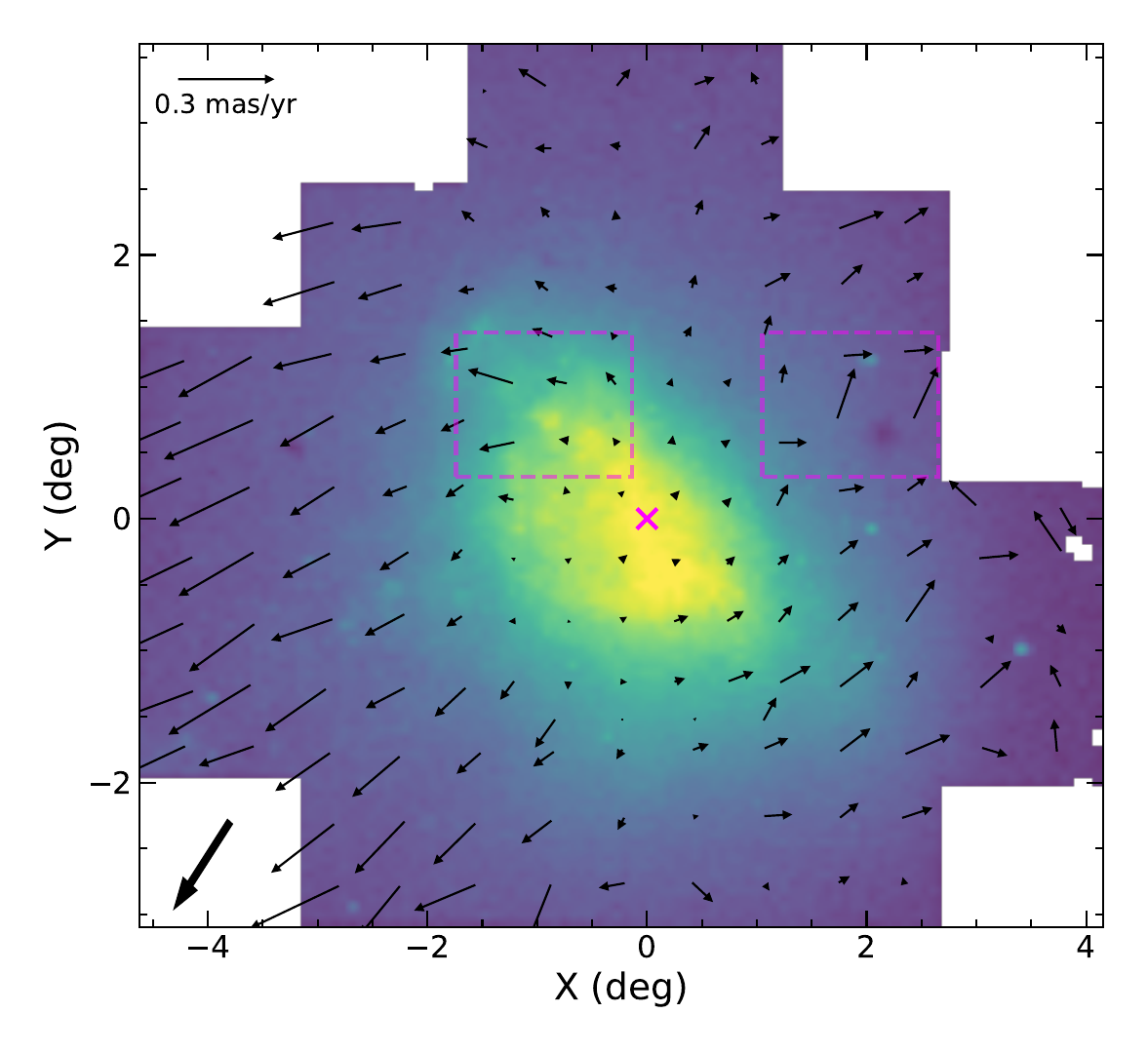}
    \hfill
    \includegraphics[width=\columnwidth]{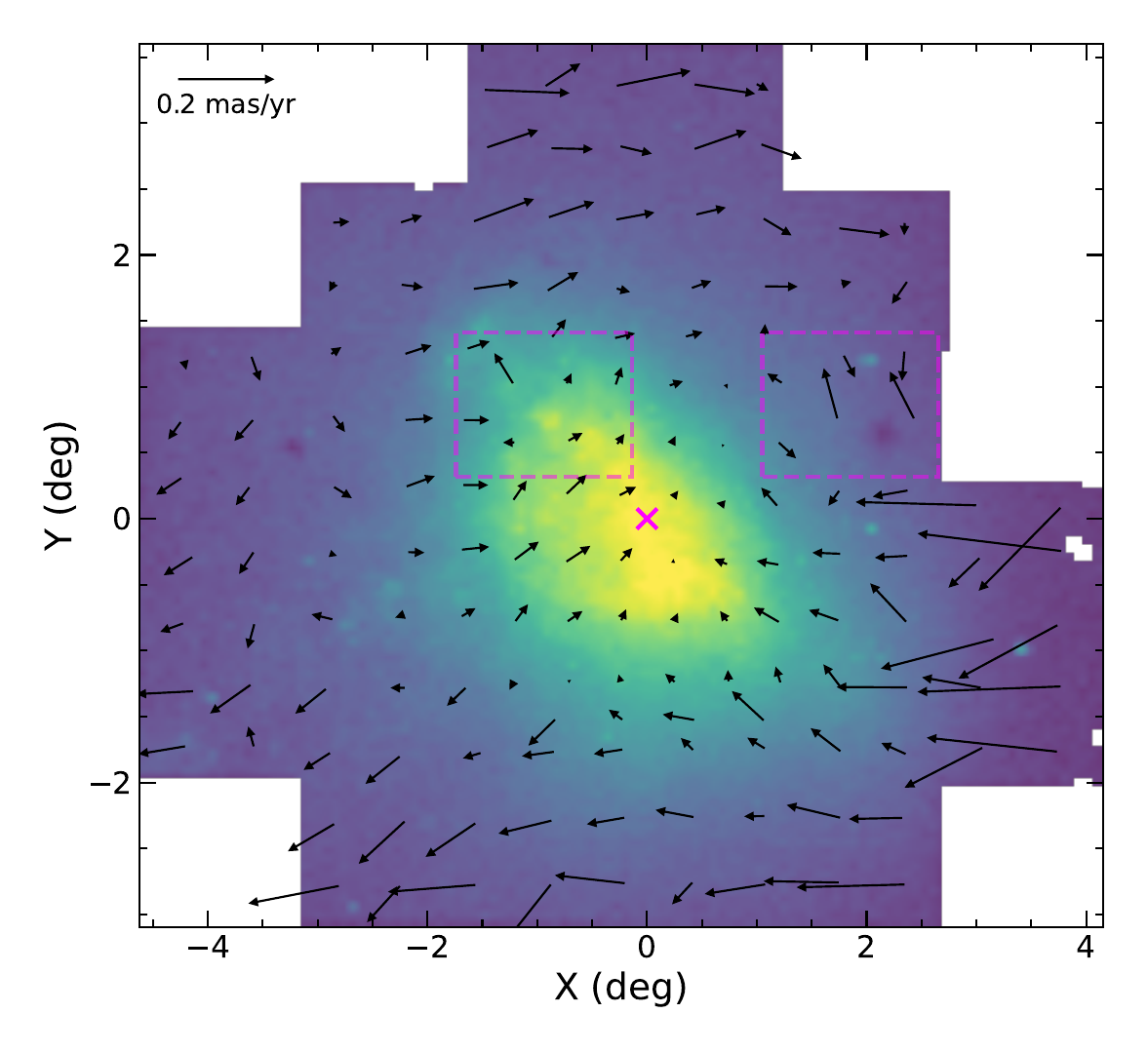}
    \caption{Residual proper motion maps projected onto the sky plane. Left: Motion of SMC stars after subtraction of the bulk systemic motion (bold arrow). Right: Motion of SMC stars after subtraction of a best-fitting linear velocity gradient from the residual motion shown in the left panel. The background shows the density of probable SMC sources in the VMC–\textit{Gaia} EDR3 cross-matched catalogue. The red dashed boxes indicate tiles SMC 5_4 and SMC 5_2 (left to right), and the optical centre is marked by a red cross. A reference vector at the top left indicates the velocity scale. East is to the left and north is at the top.}
    \label{fig:vel_map}
\end{figure*} 

For the modelling, the systemic line-of-sight velocity ($V_{\mathrm{sys}}$) and the distance to the SMC ($D_0$) were fixed to 145.6 km s$^{-1}$ \citep{2006harris} and 62.1 kpc \citep{2014graczyk}, respectively, and the dynamical centre was set to the optical centre of SMC \citep[$\alpha = 13.05$ deg, $\delta = -72.83$ deg;][]{1972vaucouleurs}, while the systemic proper motion components were kept as free parameters. The proper motion data were binned onto a 100$\times$100 spatial grid, retaining only elements containing at least 100 stars. For each grid element, covering an area of 17.2 arcminutes$^2$ on the sky, we computed the median position and proper motion, with uncertainties as the standard error of the mean. We used the Markov chain Monte Carlo (MCMC) sampling approach, implemented using the \texttt{emcee} Python package \citep{2013emcee}, to sample the posterior distribution of the model parameters by maximising the likelihood of the observed data given the model. The resulting systemic proper motions are $\mu_W$=$-0.695 \substack{+0.003 \\ -0.002}$ mas yr$^{-1}$ and $\mu_N$= $-1.247\substack{+0.003 \\ -0.002}$ mas yr$^{-1}$, in agreement with recent studies based on \textit{Gaia} DR3 data \citepalias{2021GaiaMC, 2025dhanush, 2025nakano}, suggesting a trend towards smaller values relative to earlier measurements. In Appendix \ref{sec:sys_pm}, we test the sensitivity of the SMC systemic motion to different dynamical centres for young and old populations.

\subsection{Residual proper motion maps}

In this section, we analyse the residual proper motion field in the plane of the sky. We first construct a residual velocity map by subtracting the systemic motion of the SMC from the observed proper motions. We then fit and remove a linear velocity gradient from this residual field (see top panel of Fig. \ref{fig:diver_model}), obtaining a gradient-corrected residual map that emphasises small-scale kinematic substructures. A detailed description of the adopted linear divergence model is provided in Appendix \ref{app:diver_model}. The final maps are projected onto the plane of the sky using an orthographic projection centred on the SMC as shown in Fig. \ref{fig:vel_map}. Here, tiles SMC 5$\_$2 and 5$\_$4, which have shorter time baselines, are indicated by red dashed boxes and show slightly larger residuals than adjacent regions, except for the portion of tile SMC 5$\_$4 that overlaps with the SMC gap tile. To account for strong spatial variations in stellar density, residual proper motions were binned separately in the inner (\textit{r} $\leq$ $\sim$ 2 kpc) and outer (\textit{r} > $\sim$ 2 kpc) regions. For each grid element, the median proper motion was computed using only bins containing more than 50 stars to ensure robust statistics. Residual proper motion maps were also constructed separately for different stellar populations to assess population-dependent kinematic effects (See Figs. \ref{fig:vel_map_old} and \ref{fig:vel_map_young}).

The residual velocity map reveals strong tidal expansion away from the SMC centre, with motions preferentially aligned along the south-east--north-west direction (see left panel of Fig.\ref{fig:vel_map}). In the south-east, residual motions roughly follow the SMC’s bulk motion. In the inner region, stars show streaming motions towards the east and west (see Fig. \ref{fig:vel_map_inner}), with no evidence of circular motion; velocity vectors point outwards with an average amplitude of $\sim$ 17 km s$^{-1}$ at $D_0$. This pattern is consistent with an LMC-induced tidal stretching, extending into the inner parts. While \citetalias{2025nakano} observed a similar diverging pattern in classical Cepheids, our global residual proper motion map shows a clear and coherent signature of the SMC’s tidal expansion encompassing all stellar populations.

An analysis of the gradient-corrected residual proper motion map shows that stars move towards the SMC centre in the inner regions, consistent with the central gravitational potential, while outward motions exist at the periphery, indicative of tidal forces exerted by the LMC. To test for intrinsic rotation, we fitted a modified linear velocity gradient model in which the rotational component is explicitly removed, thus allowing only anisotropic expansion and shear (see bottom panel of Fig. \ref{fig:diver_model}). Any genuine rotation would therefore appear in the residuals. The resulting velocity field is predominantly radial and shows no coherent circular pattern, indicating that the observed motions are dominated by tidal expansion rather than intrinsic rotation (see Fig. \ref{fig:vel_map_norotation}). While rotating-disk models have reported central rotation in the SMC (\citetalias{2021zivick,2025dhanush}), such models assume circular orbits and a fixed rotation axis, which can map tidal or streaming motions into an apparent rotational signal. In contrast, the absence of an antisymmetric pattern in our residual field demonstrates that the SMC’s internal kinematics are governed by tidal expansion, not coherent rotation, even in the inner region.

Another notable feature in the residual velocity map is a coherent northward motion near \textit{X} = 0.5 deg and \textit{Y} = 2.5 deg, previously interpreted as a kinematic signature of the Counter-Bridge (CB) (\citealt[hereafter D12]{2012diaz}; \citetalias{2021florian}). To examine the association of this motion with the CB, residual vectors were derived from proper motion data computed from the N-body simulation of \citetalias{2012diaz}, following the same methodology applied to the observational data. Using the distance information from the simulation, residual velocity maps were constructed for two distance bins, spanning 50–70 kpc and 70–90 kpc. The CB is expected to lie within the latter bin, at a distance of $\sim$ 85 kpc. The residual vectors in this bin exhibit a coherent north-west motion, consistent with the observed feature (see Fig. \ref{fig:CB}). Further, visual inspection of residual maps for different stellar populations shows that this motion is present only in the red giant branch (RGB) stars (see Fig. \ref{fig:vel_map_old}) and is absent in others, especially in the gradient-corrected residual velocity map, indicating pronounced population-dependent kinematics. This finding is counter-intuitive to the argument in \citetalias{2012diaz}, who suggested the CB as a complementary counterpart to the Magellanic Bridge, and should be traced by intermediate-age or young stars. Moreover, the absence of this feature in the red clump (RC) stars is unlikely to arise from the presence of a dual RC population, as \citet{2021omkumar} reports no evidence for such a population in the northern region of the SMC, using the dual RC population as evidence for line-of-sight distance bi-modality. This indicates an origin of this feature in a much older SMC interaction ($>$ 2 Gyr ago), whose imprint persists in the older RGB population \citep{2008bekki, 2023saroon} (see also Appendix \ref{app:cmd}).

Furthermore, a comparison of the residual velocity maps for the RGB and RC populations reveals that the south-east motion away from the SMC centre has a higher amplitude in the RC than in the RGB. A similar pattern is seen in the young population, comprising main-sequence and supergiant stars (see Fig.~\ref{fig:vel_map_young}). In the inner regions, the two populations show distinct behaviour: the RC exhibits a prominent south-east motion near \textit{X} = $-$1 deg, \textit{Y} = 1 deg, while the RGB moves north-west near \textit{X} = 1 deg, \textit{Y} = $-$1 deg. These differences likely reflect the varying dynamical responses of the populations: the older, dynamically hotter RGB stars exhibit less coherent residual motions, while the intermediate-age RC and young stars respond more strongly and coherently to tidal perturbations.  

In the young stellar population, no evidence of rotational motion is detected, even within the inner regions, in both the residual and gradient-corrected residual maps. The supergiant star distribution exhibits a shell-like feature in the north-east (\textit{X} = $-$1.5°, \textit{Y} = 1.25°), where stars show outward motions from the SMC centre in both maps, consistent with a tidal origin previously reported by \citet{2019martinez-delgado}. For the asymptotic giant branch (AGB) stars, which represent a small fraction of the sample, no coherent kinematic pattern is apparent, apart from a minor north-west streaming motion.

The dynamics of the SMC are highly complex, as confirmed by multiple studies, including the present work. Recent analyses \citep{2019murray,2020deleo} have highlighted that simple rotating-disk models are insufficient to capture the observed kinematics, a conclusion that is clearly supported by our results, which reveal strong signatures of disequilibrium even in the inner regions.  A natural next step in the kinematic analysis performed here is to include the third velocity component, the line-of-sight motion, for multiple stellar populations. This will be provided by the 4MOST One Thousand and One Magellanic Fields \citep[1001MC;][]{2019cioni} survey, which will enable a chemo-kinematic study of the Magellanic Clouds. 

\begin{acknowledgements}
This research was supported by the Deutsche Forschungsgemeinschaft (DFG, German Research Foundation) - Cl 213/10-1. SS acknowledges support from the Alexander von Humboldt Foundation. We thank the Cambridge Astronomy Survey Unit and the
Wide Field Astronomy Unit in Edinburgh for providing
calibrated data products under the support of the Science and
Technology Facility Council. This study is based on observations obtained with VISTA at the Paranal Observatory under programmes 179.B-2003, 099.D-0194(A), 0103.D-0161(A), 105.2042, 0109.230A, 0103.B-0783(A,B,C,D), and 109.231H. This work has made use of data from the European Space Agency (ESA) mission \textit{Gaia} (\url{https://www.cosmos.esa.int/gaia}), processed by the \textit{Gaia} Data Processing and Analysis Consortium (DPAC, \url{https://www.cosmos.esa.int/web/gaia/dpac/consortium}). This work made extensive use of several Python libraries like Astropy \citep{astropy:2013,astropy:2018, astropy:2022}, Matplotlib \citep{matplotlib}, SciPy \citep{scipy} and NumPy \citep{numpy}. We thank the anonymous referee for their constructive comments which improved the quality of the paper.
\end{acknowledgements}

\bibliographystyle{aa}
\bibliography{biblio.bib}

\twocolumn
\begin{appendix}

\section{Overview of VMC tiles}

Details of all VMC tiles used in this work are listed in Table \ref{tab:Tiles}.

\begin{table}[h!]\tiny
\setlength{\tabcolsep}{3pt}
\centering
\caption{Details of all 28 SMC tiles in the VMC survey, including the gap tile.}
\begin{tabular}{cccrccc}
\hline \hline
\noalign{\smallskip}
Tile & RA$_{\mathrm{J}2000}$ & Dec$_{\mathrm{J}2000}$ & PA & N & T$_{\mathrm{old}}$ & T$_{\mathrm{new}}$ \\ 
 & (hh:mm:ss) & (deg:mm:ss) & (deg) & & (years) & (years) \\
\noalign{\smallskip}
\hline 
\noalign{\smallskip}
SMC gap & 00:54:57.672 & $-$72:00:44.640 & +0.0002 & 12 & $-$ & 5.31 \\
SMC_2_2 & 00:21:43.920 & $-$75:12:04.320 & $-$6.7623 & 14 & 2.98 & 6.25 \\
SMC_2_3 & 00:44:35.904 & $-$75:18:13.320 & $-$1.2924 & 14 & 3.11 & 9.08 \\
SMC_2_4 & 01:07:33.864 & $-$75:15:59.760 & 4.2022 & 14 & 3.00 & 6.16 \\
SMC_2_5 & 01:30:12.624 & $-$75:05:27.600 & 9.6169 & 14 & 3.16 & 9.08 \\
SMC_3_1 & 00:02:39.912 & $-$73:53:31.920 & $-$11.3123 & 14 & 1.86 & 6.96 \\
SMC_3_2 & 00:23:35.544 & $-$74:06:57.240 & $-$6.3137 & 14 & 3.14 & 6.18 \\
SMC_3_3 & 00:44:55.896 & $-$74:12:42.120 & $-$1.212 & 14 & 1.09 & 8.04 \\
SMC_3_4 & 01:06:21.120 & $-$74:10:38.640 & 3.9099 & 14 & 2.94 & 6.17 \\
SMC_3_5 & 01:27:30.816 & $-$74:00:49.320 & 8.9671 & 14 & 1.19 & 8.06 \\
SMC_3_6 & 01:48:06.120 & $-$73:43:28.200 & 13.8809 & 14 & 2.82 & 6.88 \\
SMC_4_1 & 00:05:33.864 & $-$72:49:12.000 & $-$10.6178 & 14 & 2.83 & 6.07 \\
SMC_4_2 & 00:25:14.088 & $-$73:01:47.640 & $-$5.9198 & 14 & 1.68 & 6.91 \\
SMC_4_3 & 00:45:14.688 & $-$73:07:11.280 & $-$1.1369 & 14 & 1.92 & 7.90 \\
SMC_4_4 & 01:05:19.272 & $-$73:05:15.360 & 3.6627 & 14 & 1.80 & 6.87 \\
SMC_4_5 & 01:25:11.016 & $-$72:56:02.040 & 8.4087 & 14 & 2.24 & 7.94 \\
SMC_4_6 & 01:44:34.512 & $-$72:39:44.640 & 13.0368 & 14 & 2.83 & 8.95 \\
SMC_5_2 & 00:26:41.688 & $-$71:56:35.880 & $-$5.5717 & 13 & 1.42 & $-$ \\
SMC_5_3 & 00:44:49.032 & $-$72:01:36.120 & $-$1.2392 & 14 & 1.87 & 6.95 \\
SMC_5_4 & 01:04:26.112 & $-$71:59:51.000 & 3.4514 & 13 & 2.00 & $-$ \\
SMC_5_5 & 01:23:04.944 & $-$71:51:47.880 & 7.6718 & 14 & 2.83 & 9.00 \\
SMC_5_6 & 01:41:28.800 & $-$71:35:47.040 & 12.3004 & 14 & 2.95 & 10.92 \\
SMC_6_2 & 00:27:39.960 & $-$70:51:12.600 & $-$5.3423 & 14 & 2.75 & 9.00 \\
SMC_6_3 & 00:45:48.768 & $-$70:56:08.160 & $-$1.0016 & 14 & 2.21 & 11.10 \\
SMC_6_4 & 01:03:49.944 & $-$70:53:34.440 & 3.1075 & 14 & 2.75 & 9.03 \\
SMC_6_5 & 01:21:22.488 & $-$70:46:10.920 & 7.5039 & 14 & 2.12 & 11.04 \\
SMC_7_3 & 00:46:04.728 & $-$69:50:38.040 & $-$0.9389 & 14 & 2.84 & 8.90 \\
SMC_7_4 & 01:03:00.480 & $-$69:48:58.320 & 3.1144 & 14 & 2.74 & 9.89 \\
\noalign{\smallskip}
\hline
\end{tabular}
\label{tab:Tiles}
\tablefoot{\tiny{The RA and Dec columns give the central coordinates of the tiles. The columns PA, N, T$_{\mathrm{old}}$ and T$_{\mathrm{new}}$ denote the position angle measured from North to East, the number of epochs and the old and new time baselines, respectively. The proper motion tables for all VMC tiles will be included in the final data release of the VMC survey (DR7), which is available through the VMC ESO Phase3 collection (\url{https://doi.eso.org/10.18727/archive/64/}). For further details regarding the data release, see \citet{2025cioni}.}}
\end{table}

\section{Near-infrared colour-magnitude diagram for SMC sources}
\label{app:cmd}

\begin{figure}[htbp]
    \centering
    \includegraphics[width=0.9\columnwidth]{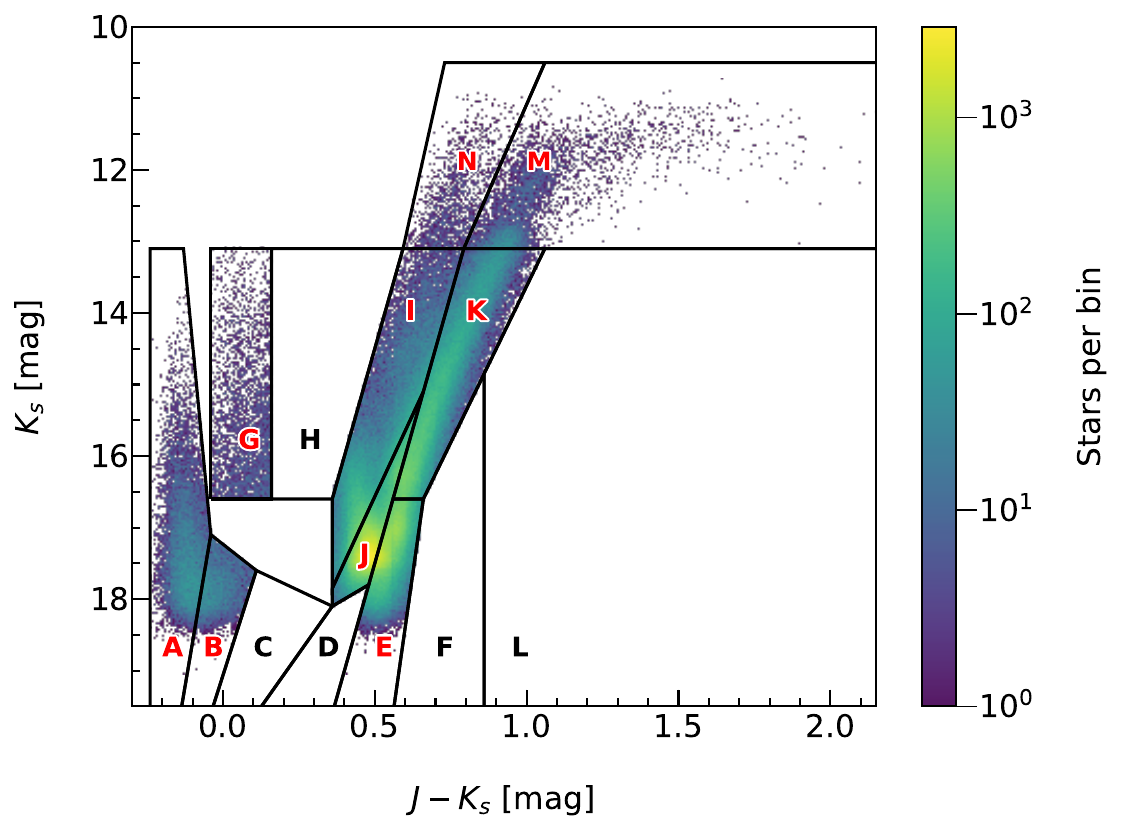}
    \caption{Near-infrared CMD of probable SMC sources in the VMC dataset. The segments dominated by SMC sources are highlighted in red, overlaid on the stellar density distribution.}
    \label{fig:cmd}
\end{figure}

\begin{figure}[htbp]
    \centering
    \includegraphics[width=0.9\columnwidth]{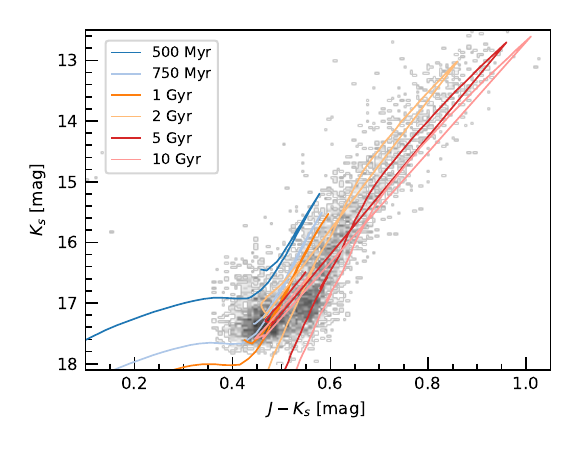}
    \caption{CMD with isochrones of different ages overlaid for stars at the location of the Counter-Bridge motion.}
    \label{fig:cmd_iso}
\end{figure}

The near-infrared CMD is employed to distinguish between probable SMC sources from MW stars and to separate the different stellar populations. The CMD is divided into different segments following the selection criteria of \citet{2018dalal}. The segments were derived based on a detailed study of the star formation history of the galaxy \citep{2012rubele}. There are 14 segments corresponding to various stellar populations, including young main-sequence stars (A, B, and C), intermediate-age main-sequence and sub-giant stars (D), faint and bright RGB stars (E, K), RC stars (J), supergiants (G, H, I, N), AGB stars (M), Milky Way foreground stars (F), and background galaxies (L). For this work, we selected segments A ($\sim$ 20 Myr), B ($\sim$ 141 Myr), E ($\sim$ 4.46 Gyr), G ($\sim$ 112 Myr), I ($\sim$ 512 Myr), J ($\sim$ 4.07 Gyr), K ($\sim$ 4.26 Gyr), M ($\sim$ 2.45 Gyr), and N ($\sim$ 85.1 Myr)\footnote{The median age for segment N was computed in this work}; where the values in parentheses denote the median ages of the corresponding stellar populations \citep{2019dalal, 2023dalal}. These segments are dominated by SMC sources (see Fig.~\ref{fig:cmd}). Although RC stars are generally younger than RGB stars, the apparent similarity in their ages arises from RGB contamination in the J segment. This is illustrated in the CMD for stars in the northern region, where a potential Counter-Bridge signature is observed (see Fig. \ref{fig:cmd_iso}), with PARSEC isochrones \citep{2012bressan} overlaid, assuming a metallicity of $-$1.0 dex for the SMC \citep{2023mucciarelli}. In this region, the RC is dominated by stars aged between 750 Myr and 1 Gyr, while the RGB consists of stars older than 1 Gyr, consistent with our interpretation that the observed signature may result from an earlier SMC interaction (>2 Gyr ago).
 
\section{Confusion in VMC proper motion catalogue}
\label{app:confusion_limit}

\begin{figure}[htbp]
    \centering
    \includegraphics[width=0.93\columnwidth]{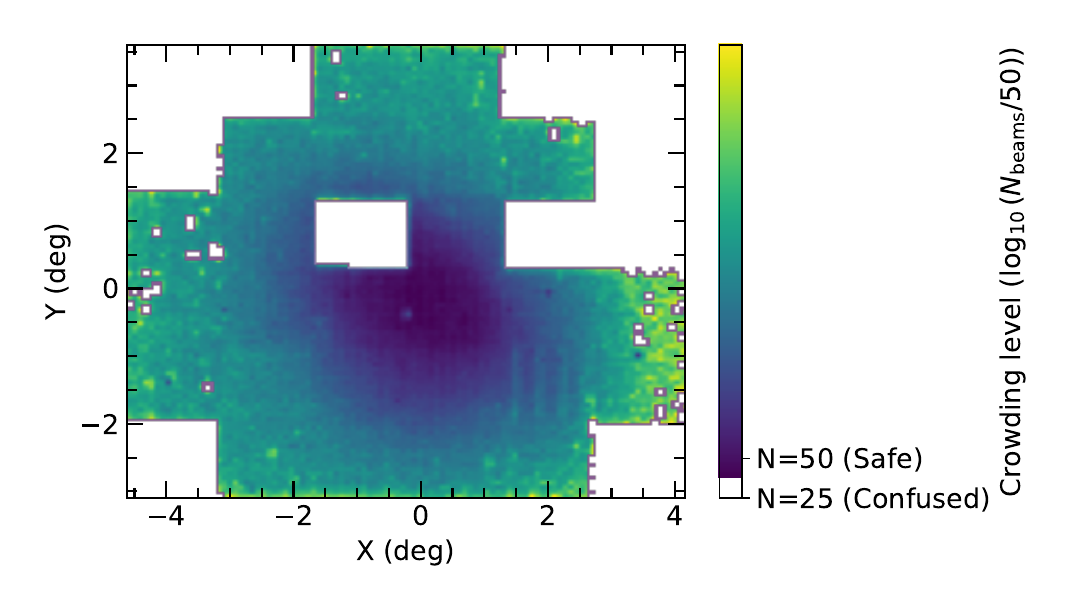}
    \includegraphics[width=0.93\columnwidth]{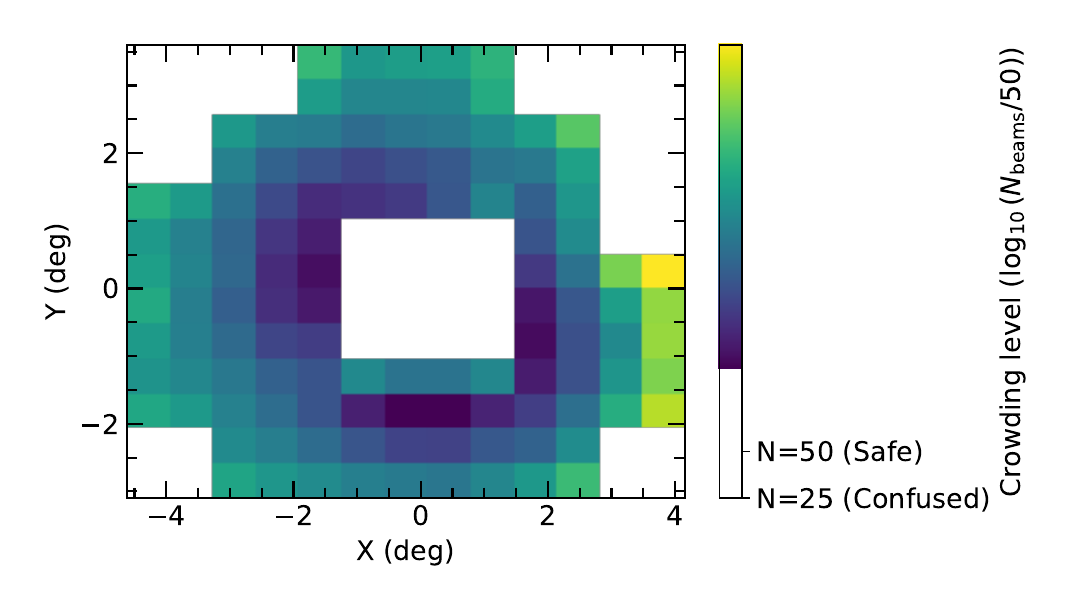} 
    \includegraphics[width=0.93\columnwidth]{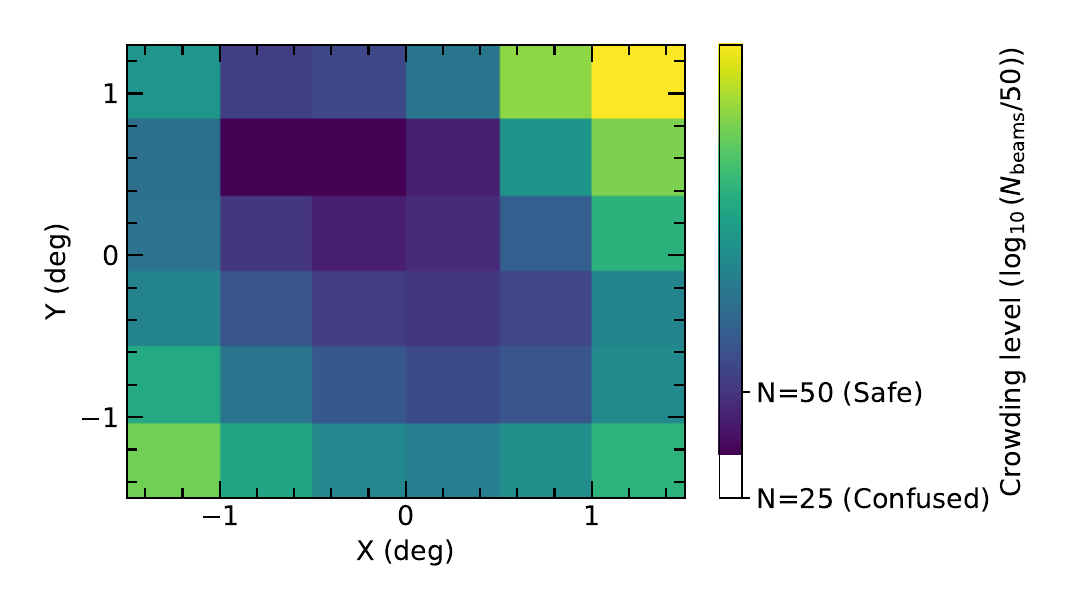} 
    \caption{Confusion maps for different spatial binning. Top: 100$\times$100 binning used for systemic motion, with tiles SMC 5$\_$2 and SMC 5$\_$4 removed. Middle: 13$\times$13 binning applied to the outer region for the proper motion map. Bottom: 6$\times$6 binning applied to the inner region binning for the proper motion map. The threshold number of beams (\textit{N} = 50) and the confused value (\textit{N} = 25) are indicated on the colour bar.}
    \label{fig:confusion_map}
\end{figure}

Confusion is defined as the loss of reliable source separation that occurs when multiple sources occupy a single resolution element (PSF or beam) and is typically quantified by the number of beams per source within a given area \citep{1974condon}. Here, we evaluate the impact of source confusion in the VMC proper motion catalogue by computing the number of beams per source for different spatial binning. The \textit{K}$_{\mathrm{s}}$-band seeing of 0.9 arcsecond \citep{2025cioni} is used as the effective resolution element. To determine the systemic motion of the SMC, we used a 100$\times$100 spatial binning, and restricted the analysis to tiles with a time baseline longer than 6 years. Two different binning schemes were applied for the inner and outer regions in the construction of the residual proper motion maps. The spatial extent of the inner region was defined as spanning from $-1.5$ deg to $1.5$ deg in the \textit{x} direction and $-1.5$ deg to $1.3$ deg in the \textit{y} direction; all remaining areas are classified as the outer region. The resulting confusion maps for all binning configurations are shown in Fig. \ref{fig:confusion_map}. We chose a conservative confusion threshold of 50 beams per source as a safe value, while fewer than 25 beams per source was considered confused. We see that the central region of the 100$\times$100 grid and a few bins of the inner region are below the safe value. However, even in this case, the values remain above 25 beams per source, commonly taken as the critical limit for significant confusion effects, especially for near-infrared wavelengths.

\section{Geometric framework for modelling the systemic proper motion}
\label{app:model}

\begin{figure}[htbp]
    \centering
    \includegraphics[width=0.85\columnwidth]{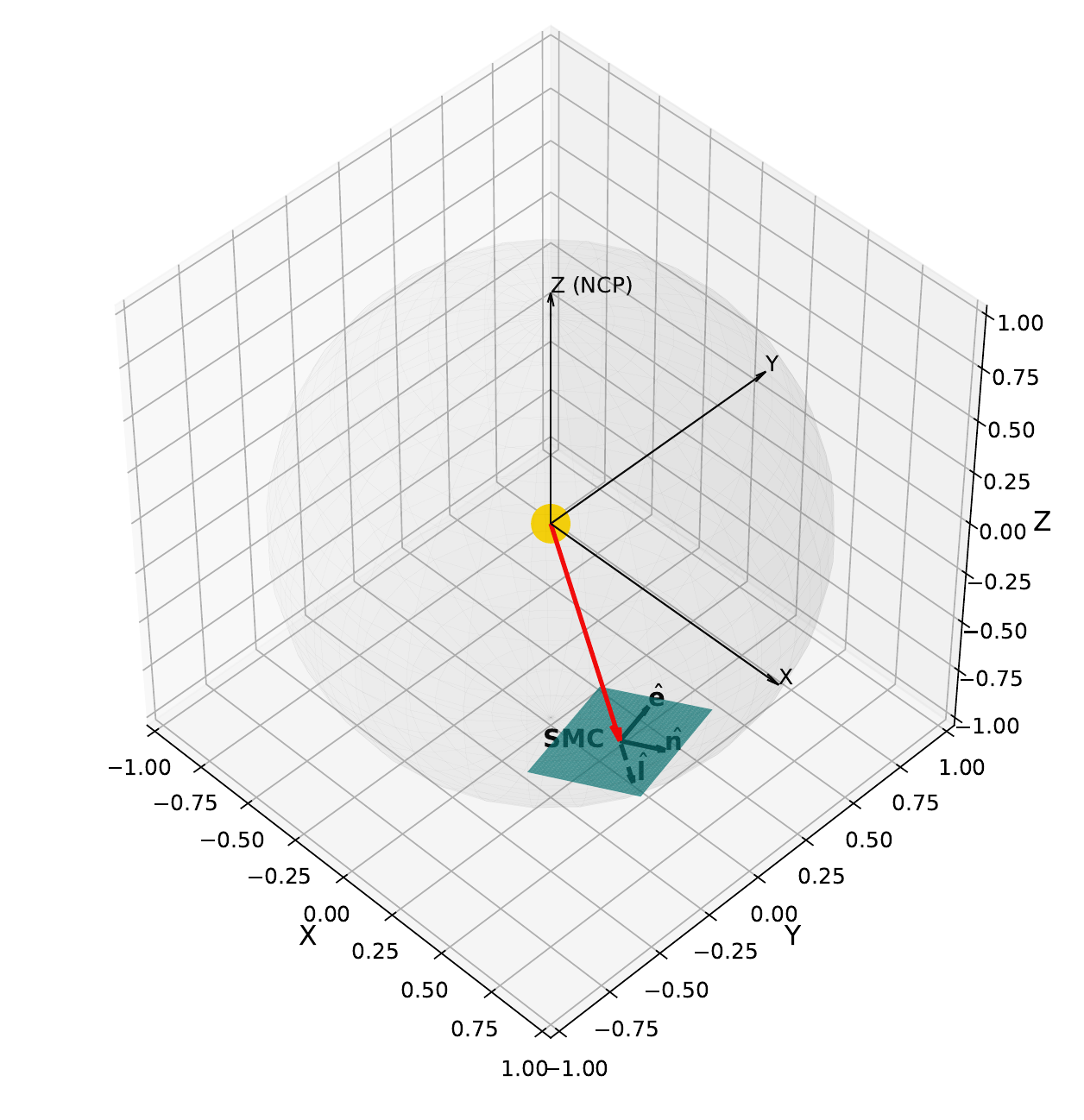}
    \caption{3D view of the coordinate systems. The yellow dot marks the Sun while the red arrow represents the positional vector $\boldsymbol{r}$ to the SMC. The tangent plane at the dynamical centre of SMC is shown in cyan.}
    \label{fig:geo_coor}
\end{figure}

The galaxy is represented by a rigid body without rotation moving in a three-dimensional (3D) space. We define a 3D Cartesian coordinate system centred on the Sun, with the \textit{XY}-plane aligned with the celestial equator (Dec = 0 deg) and the \textit{Z}-axis pointing towards the north celestial pole. The position of the SMC is described by the vector $\boldsymbol{r}$. A tangent plane is defined at the position of the SMC dynamical centre, spanned by the orthonormal basis vectors ($\hat{\boldsymbol{e}}$, $\hat{\boldsymbol{n}}$, $\hat{\boldsymbol{l}}$), corresponding to the directions of increasing RA (east), Dec (north), and the line of sight, respectively (see Fig. \ref{fig:geo_coor}). The vectors $\hat{\boldsymbol{e}}$, $\hat{\boldsymbol{n}}$ are obtained from the normalised partial derivatives of $\boldsymbol{r}$  with respect to $\alpha$ and $\delta$, while $\hat{\boldsymbol{l}}$ is the normalised position vector. The velocity of the SMC dynamical centre is expressed as:
\begin{equation}
    \boldsymbol{v}_{0} = V_{\mathrm{sys}} \hat{\boldsymbol{l}_0}(\alpha_0, \delta_0) + D_0\mu_{N,0} \hat{\boldsymbol{n}_0}(\alpha_0, \delta_0) - D_0\mu_{W,0} \hat{\boldsymbol{e}_0}(\alpha_0, \delta_0)
    \footnotemark
\end{equation}
\footnotetext{The conversion factor from mas yr$^{-1}$ to km s$^{-1}$ is not shown in the equation, but should be included in calculations.}

\noindent where $V_{\mathrm{sys}}$ and $D_0$ are the systemic line-of-sight velocity and distance to the SMC and $\mu_{W,0}$ and $\mu_{N,0}$ are the systemic proper motion components. We model the projection of the systemic motion for each star (\textit{i}) across the full spatial extent, taking into account the perspective effect due to line-of-sight motion as: 

\begin{equation}
    \mu_{W,\mathrm{mod},i} = -\frac{\boldsymbol{v}_{0} \cdot \hat{\boldsymbol{e_i}}}{D_0}, \quad
    \mu_{N,\mathrm{mod},i} = \frac{\boldsymbol{v}_{0} \cdot \hat{\boldsymbol{n_i}}}{D_0}
\end{equation}

In this modelling, we assume a fixed line-of-sight distance ($D_0$ = 62.1 kpc) and velocity ($V_{\mathrm{sys}}$ = 145.6 km/s) for the SMC. To quantify the sensitivity of these assumptions, we vary $D_0$ by $\pm$ 30 kpc and $V_{\mathrm{sys}}$ by $\pm$ 20 km/s at an angular separation of 4 deg from the galactic centre. We find that the resulting change in proper motion is $\Delta \mu$ $\sim \pm $ 0.0048 mas yr$^{-1}$ for $\Delta$$V_{\mathrm{sys}}$, and $\Delta \mu$ $\sim$ $-$0.011 to $+$0.032 mas yr$^{-1}$ for $\Delta D$. These variations are small compared to the measured residual proper motions. The effect of $\Delta D$ is roughly an order of magnitude smaller, especially in the eastern region, while $V_{\mathrm{sys}}$ produces changes two orders of magnitude smaller. Therefore, uncertainties in $D_0$ and $V_{\mathrm{sys}}$ have negligible impact on the residual proper motion maps.

\section{Systemic proper motion for different dynamical centres}
\label{sec:sys_pm}
We tested three literature dynamical centres for the SMC to assess systematic effects arising from the choice of centre: the optical centre \citep[$\alpha = 13.05$ deg, $\delta = -72.83$ deg;][]{1972vaucouleurs}, the H \textsc{i} dynamical centre \citep[$\alpha = 16.26$ deg, $\delta = -72.42$ deg;][]{2004stanimiorvic}, and the Cepheid-based centre \citep[$\alpha = 12.54$ deg, $\delta = -73.11$ deg;][]{2017ripepi}. The inferred systemic proper motions in $\mathrm{mas \,yr^{-1}}$ (see Table~\ref{tab:param}) depend on the adopted centre: the optical and Cepheid centres yield consistent results, whereas the H \textsc{i} centre shows a significant offset of $\sim$0.06–0.07 $\mathrm{mas_\ yr^{-1}}$ in $\mu_W$ and $\sim$0.03–0.04 $\mathrm{mas_\ yr^{-1}}$ in $\mu_N$. 

\begin{table*}[h]\tiny
\centering
\caption{Systemic motion and their associated uncertainties obtained from the MCMC sampling for different dynamical centres.}
\begin{tabular}{lcccccc}
\hline\hline
\noalign{\smallskip}
Dynamical Centre \\
\tiny{(Epoch J2000)}
& \multicolumn{2}{c}{Entire Sample \tiny{($\mathrm{mas\,yr^{-1}}$)}} & \multicolumn{2}{c}{Old Population \tiny{($\mathrm{mas\,yr^{-1}}$)}} & \multicolumn{2}{c}{Young Population \tiny{($\mathrm{mas\,yr^{-1}}$)}} \\
\cmidrule(lr){2-3}
\cmidrule(lr){4-5}
\cmidrule(lr){6-7}
 & $\mu_{W,0}$  & $\mu_{N,0}$
 & $\mu_{W,0}$ & $\mu_{N,0}$
 & $\mu_{W,0}$ & $\mu_{N,0}$ \\
\noalign{\smallskip}
\hline
\noalign{\smallskip}
Optical Centre \\
$\alpha = 13.05$ deg, $\delta = -72.83$ deg & $-0.695 \substack{+0.003 \\ -0.002}$ & $-1.247\substack{+0.003 \\ -0.002}$ & $-0.694 \substack{+0.003 \\ -0.002}$ & $-1.245\substack{+0.004 \\ -0.002}$ & $-0.679 \substack{+0.025 \\ -0.012}$ & $-1.218\substack{+0.025 \\ -0.012}$ \\
\noalign{\smallskip}
Cepheid Centre \\
$\alpha = 12.54$ deg, $\delta = -73.11$ deg & $-0.685 \substack{+0.003 \\ -0.002}$ & $-1.250\substack{+0.003 \\ -0.002}$ & $-0.685 \substack{+0.003 \\ -0.002}$ & $-1.249\substack{+0.004 \\ -0.002}$ & $-0.670 \substack{+0.026 \\ -0.012}$ & $-1.222\substack{+0.026 \\ -0.012}$ \\
\noalign{\smallskip}
{H\,\sc{i}} gas Centre \\
$\alpha = 16.26$ deg, $\delta = -72.42$ deg & $-0.753 \substack{+0.003 \\ -0.002}$ & $-1.212\substack{+0.003 \\ -0.002}$ & $-0.751 \substack{+0.003 \\ -0.002}$ & $-1.210\substack{+0.004 \\ -0.002}$ & $-0.735 \substack{+0.025 \\ -0.012}$ & $-1.184\substack{+0.026 \\ -0.012}$ \\
\noalign{\smallskip}
\hline
\label{tab:param}
\end{tabular}
\tablefoot{The systemic motion is measured in $\mathrm{mas \,yr^{-1}}$.}
\end{table*}

To investigate population-dependent effects, we further separated the sample into old ($\gtrsim$1 Gyr; E, K, M, J) and young ($\lesssim$0.5 Gyr; A, B, G, N) stars based on their near-infrared CMD positions (see Fig. \ref{fig:cmd}). For all centres, the young population shows systematically lower systemic motions, though these differences remain consistent with zero within the uncertainties, reflecting the larger errors associated with its lower stellar density. 

\section{Anisotropic divergence model}
\label{app:diver_model}
An anisotropic divergence model is used for modelling the tidal expansion of the SMC induced by the interaction with the LMC. The model is formulated as a linear velocity gradient field, allowing for anisotropic expansion as well as shear and rotational components:

\begin{equation}
\boldsymbol{v} = \mathbf{G}\,\boldsymbol{r}, 
\qquad 
\mathbf{G} =
\begin{pmatrix}
a & b \\
c & d
\end{pmatrix},
\qquad
\boldsymbol{r} =
\begin{pmatrix}
x - x_0 \\
y - y_0
\end{pmatrix}
\end{equation}

\begin{figure}
    \centering
    \includegraphics[width=0.7\columnwidth]{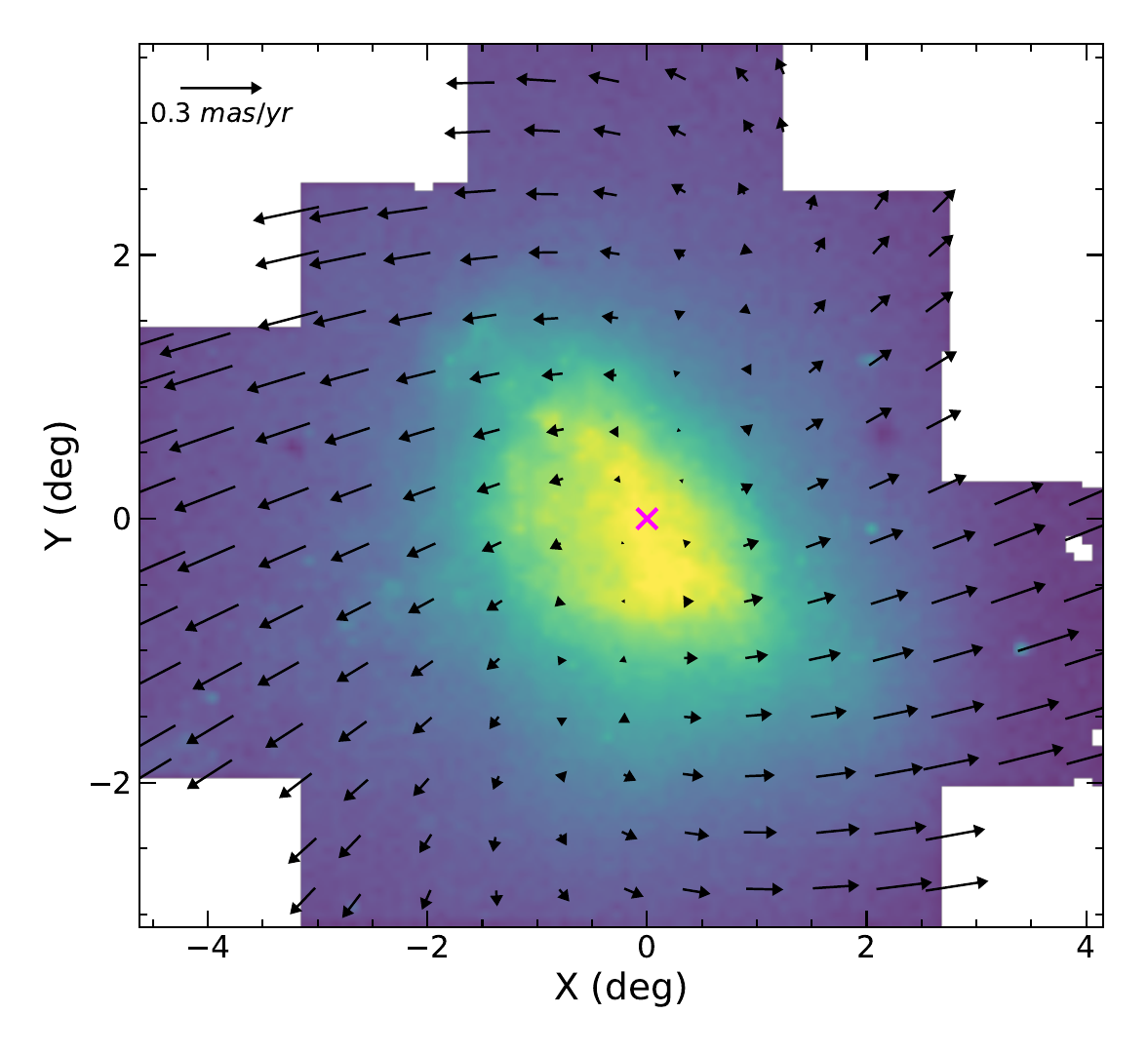}
    \hfill
    \includegraphics[width=0.7\columnwidth]{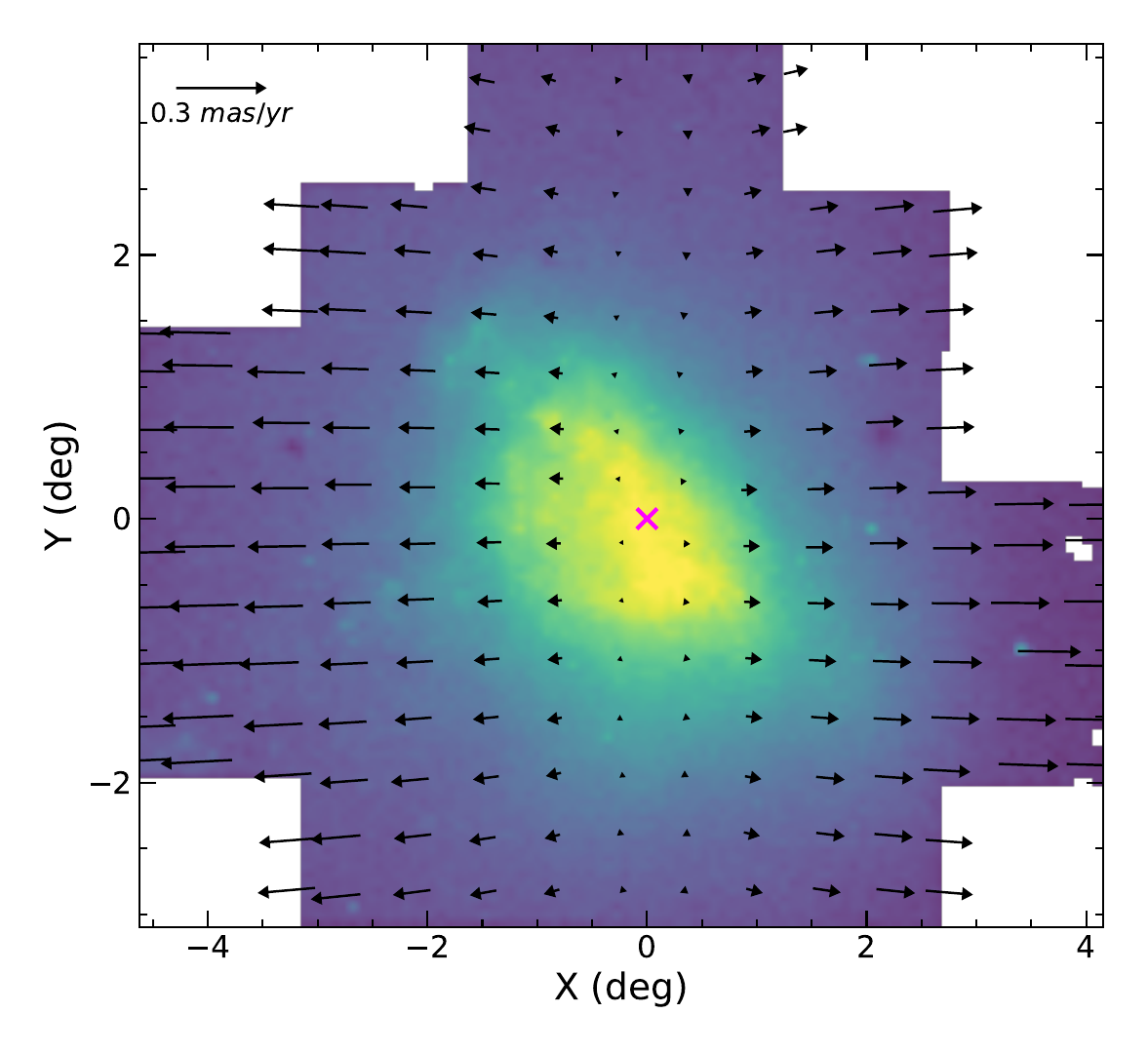}
    \caption{Same as in Fig. \ref{fig:vel_map}, but here the arrows represent the best-fitting linear velocity gradients. Top: Diverging model, including the rotation component. Bottom: Linear divergence model without rotation.}
    \label{fig:diver_model}
\end{figure}

The matrix \textbf{G} captures the anisotropic divergence in \textit{x} and \textit{y} directions, as well as any associated shear or rotational components. The diagonal terms \textit{a} and \textit{d} quantify the expansion or contraction in the x and y directions, respectively, while the off-diagonal terms \textit{b} and \textit{c} capture the shear and rotational motions. The velocity field reduces to pure solid body rotation when $a=d=0$ and $b=-c$, while configurations with $b=c$ contain no rotational component and represent pure anisotropic expansion with shear. The vector \textbf{r} represents the position of each star (\textit{x}, \textit{y}) with respect to the dynamical centre of SMC ($x_0$, $y_0$). The best-fitting divergence model for the SMC stars is shown in Figure \ref{fig:diver_model}.

The best-fitting parameters of the divergence model are \textit{a} = 0.0581, \textit{b} = $-$0.0295, \textit{c} = 0.0240 and \textit{d} = 0.0088, with uncertainties of 0.0003, 0.0011, 0.0007, and 0.0007, respectively. The relatively large value of \textit{a} compared to \textit{d} shows that the dominant velocity gradient is along the \textit{x} direction. The off-diagonal terms satisfy \textit{b} $\sim$ $-$ \textit{c}, which suggests the presence of a small curl in the velocity field. Such a curl can occur from the curvature in the tidal flow rather than disk rotation.

\begin{figure}[htbp]
    \centering
    \includegraphics[width=0.7\columnwidth]{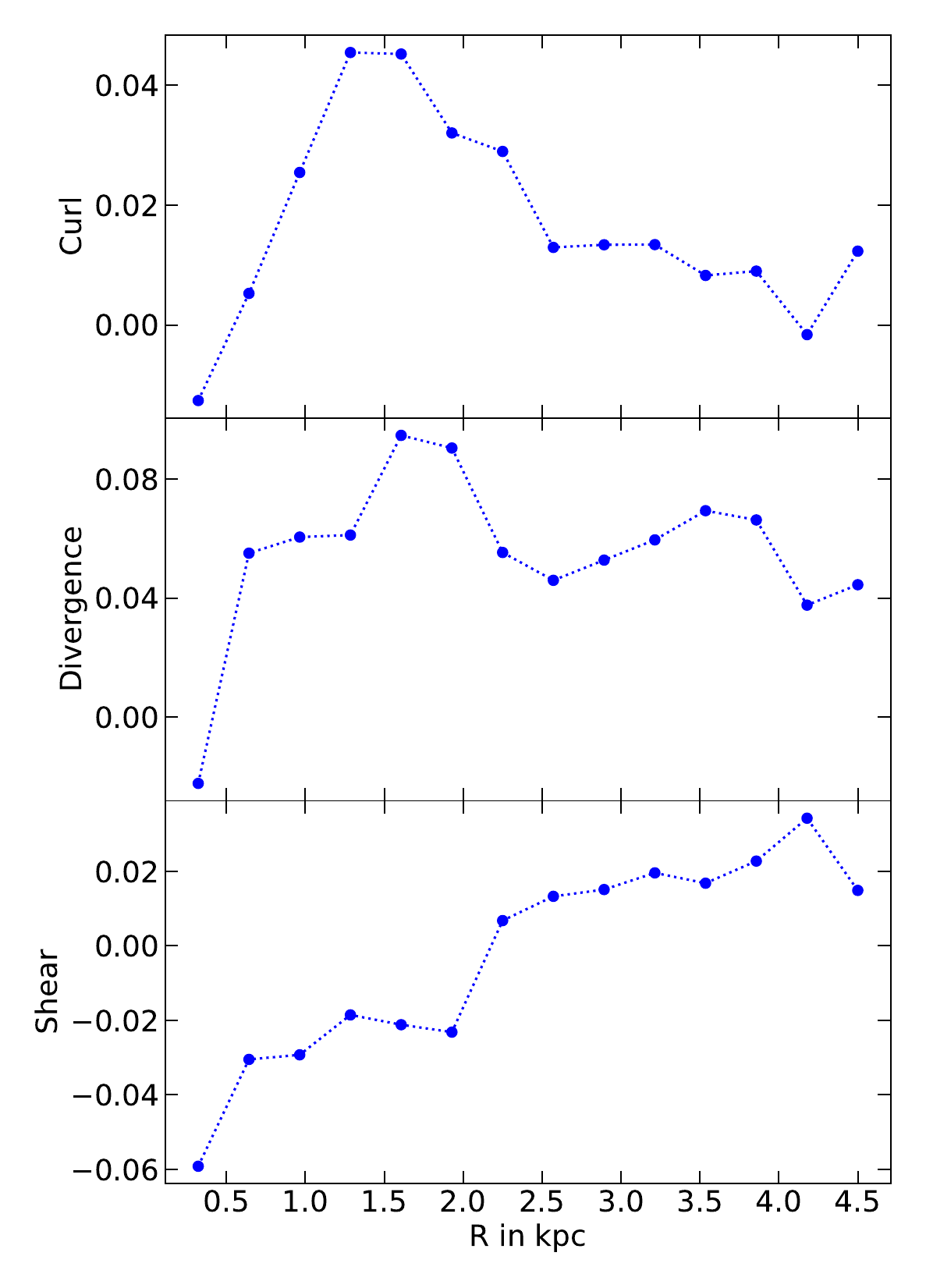}
    \caption{Different kinematic properties of the velocity gradient tensor as a function of radius. The top, middle and bottom panels show the curl, divergence and shear, respectively.}
    \label{fig:tensor}
\end{figure}

To distinguish between these properties, we examined the radial behaviour of the kinematic quantities derived from the velocity gradient tensor \textbf{G}: the divergence (\textit{a}+\textit{d}), curl ((\textit{c}$-$\textit{b})/2) and the shear ((\textit{b}$+$\textit{c})/2), as shown in Fig. \ref{fig:tensor}. The curl increases with radius up to $\sim$ 1.5 kpc and subsequently declines with a small increase after 3.5 kpc. This localised peak may reflect curvature in the tidally distorted velocity field rather than global disk rotation, which would yield a smoother radial decline. However, additional kinematic effects, which are not explicitly tested here, may contribute to this non-axisymmetric signal. Interestingly, the divergence also peaks at $\sim$ 1.5 kpc, indicating that the stellar motions experience the strongest expansion at this radius, consistent with a region of maximum tidal influence. The shear increases with radius, consistent with tidal stretching of the SMC.

\section{Residual velocity maps }

\begin{figure}[htbp]
    \centering
    \includegraphics[width=0.7\columnwidth]{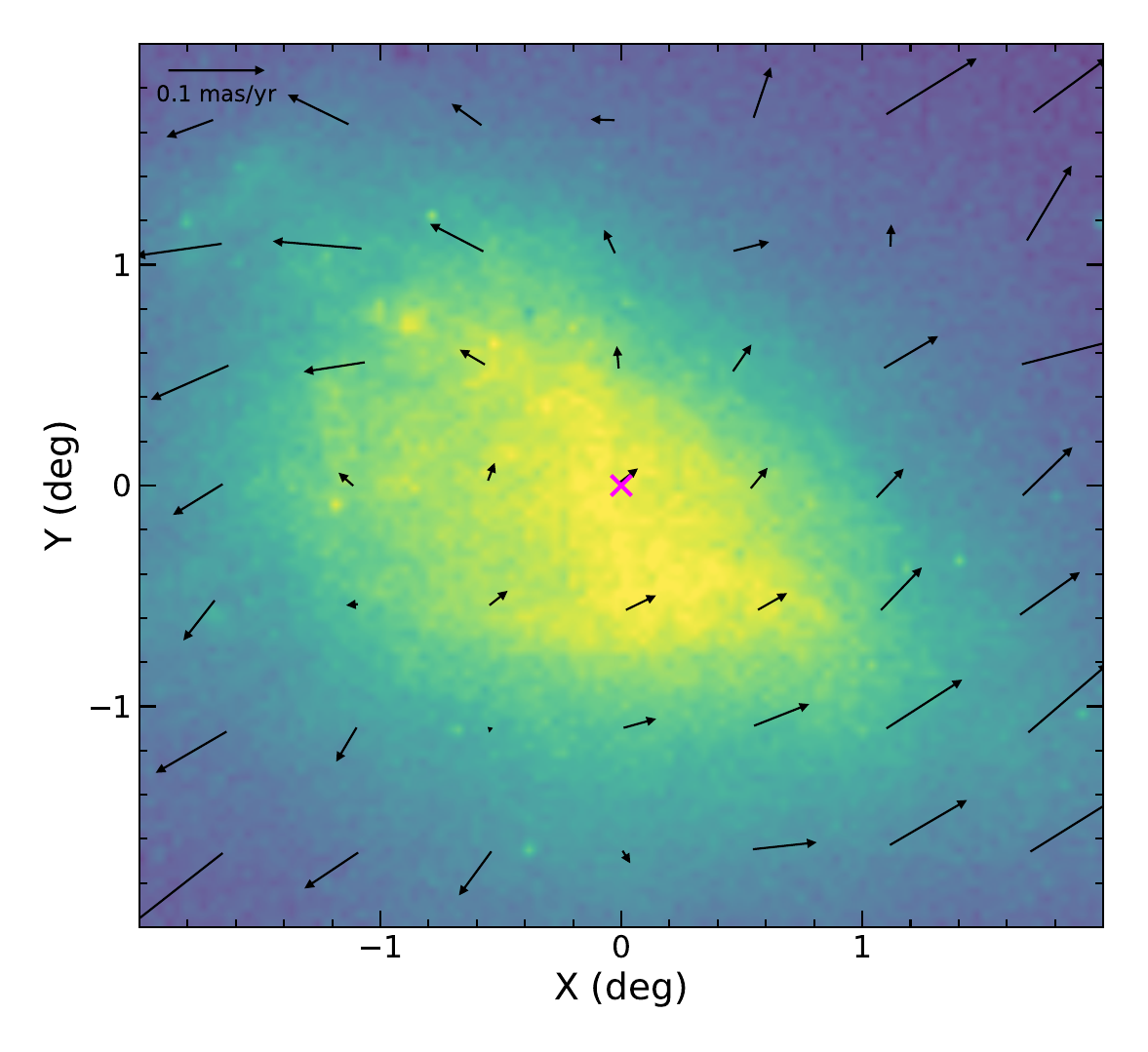}
    \caption{Zoom-in view of the inner 2 kpc region of the left panel in Fig. \ref{fig:vel_map}.}
    \label{fig:vel_map_inner}
\end{figure}

\begin{figure}[htbp]
    \centering
    \includegraphics[width=0.7\columnwidth]{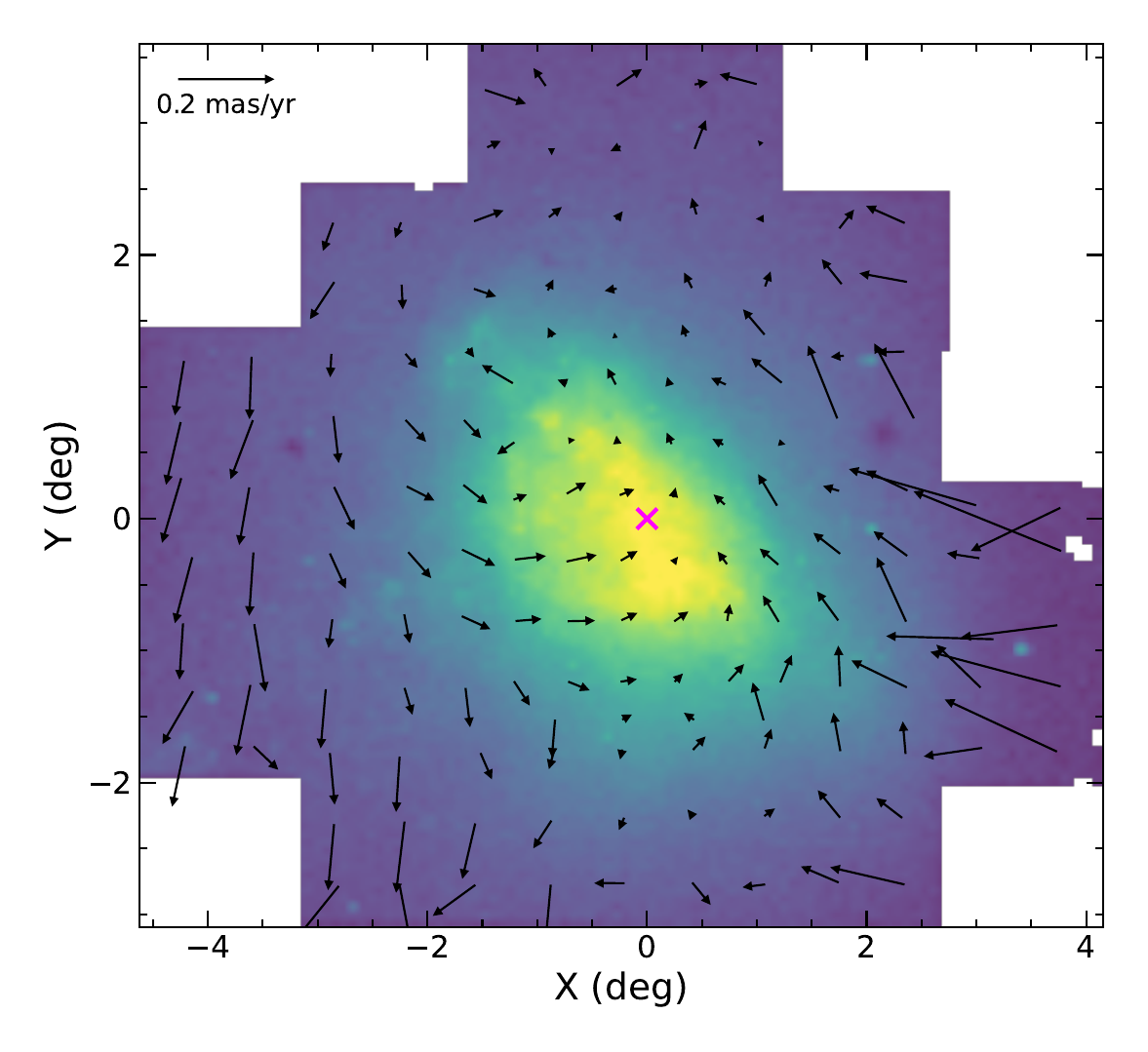}
    \caption{Same as Fig. \ref{fig:vel_map}, but showing the gradient-corrected residuals including the rotation component.}
    \label{fig:vel_map_norotation}
\end{figure}

\begin{figure}[htbp]
    \centering
    \includegraphics[width=0.8\columnwidth]{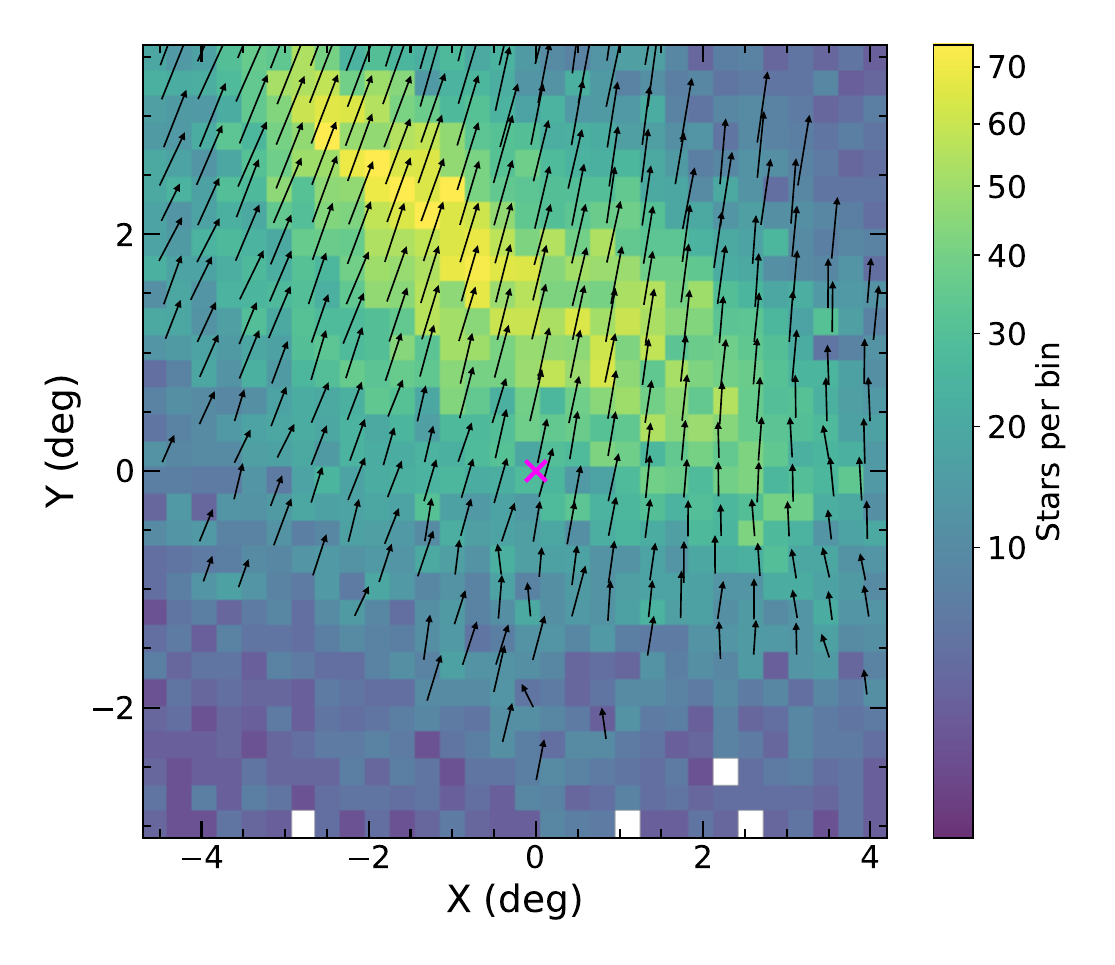}
    \caption{Residual velocity map for the distance bin covering 70--90 kpc, including the Counter-Bridge located at $\sim$ 85 kpc.}
    \label{fig:CB}
\end{figure}

The residual velocity map for the inner 2 kpc region of the SMC is shown in Fig. \ref{fig:vel_map_inner}. The diverging motion along the east-west direction is clearly visible, indicating a strong tidal force from the LMC. In Fig. \ref{fig:vel_map_norotation}, we show the gradient-corrected residual velocity map where the linear velocity gradient without modelling the rotation component was adopted to see any residual rotation motion, especially in the inner 2 kpc region. Finally, the residual velocity map for the CB region, created using proper motion data from \citetalias{2012diaz} simulation, is shown in Fig. \ref{fig:CB}. The kinematic signature of the CB towards the north-west direction is clearly evident. 

\section{Residual velocity map for different stellar populations}

\begin{figure*}[htbp]
    \centering
    \includegraphics[width=0.8\textwidth]{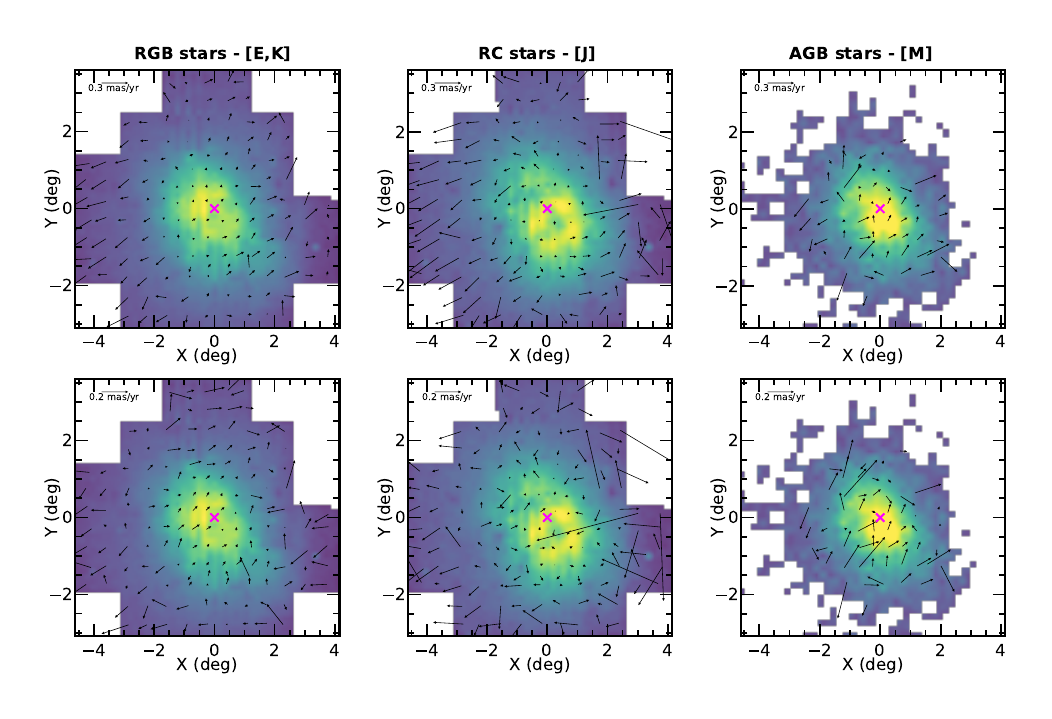}
    \caption{Velocity maps for RGB, RC and AGB stars with CMD segments. Top: Residual velocities after subtracting the systemic motion. Bottom: Gradient-corrected residual velocities.}
    \label{fig:vel_map_old}
\end{figure*}

\begin{figure*}[htbp]
    \centering
    \includegraphics[width=0.6\textwidth]{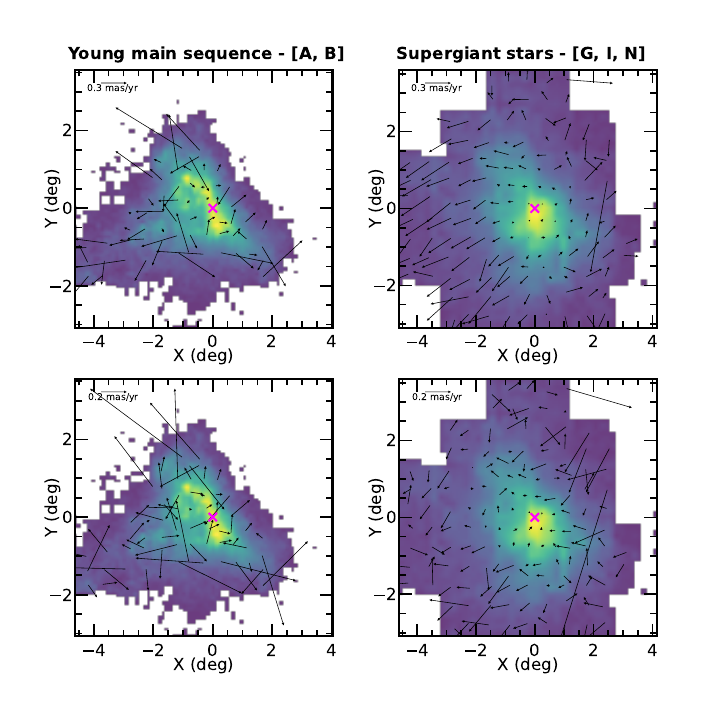}
    \caption{Same as Fig. \ref{fig:vel_map_old}, but for young stellar population.}
    \label{fig:vel_map_young}
\end{figure*}

The residual and the gradient-corrected residual velocity maps for different stellar populations are shown in Figures \ref{fig:vel_map_old}, \ref{fig:vel_map_young}. 

\end{appendix}

\end{document}